\documentclass[a4paper]{jpconf}
\usepackage{graphicx}
\newcommand{\be}{\begin{eqnarray}}
\newcommand{\ee}{\end{eqnarray}}
\begin{document}
\title{A Comment on Conical Flow \\ Induced by Heavy-Quark Jets}

\author{ F. Antinori
and ~E.V. Shuryak}

\address{ Istituto Nazionale di Fisica Nucleare and Dipartimento di Fisica Galileo Galilei, I-35131 Padova, Italy\\
Department of Physics and Astronomy\\ State University of New York,
     Stony Brook, NY 11794-3800}

\ead{federico.antinori@pd.infn.it; shuryak@tonic.physics.sunysb.edu}

\begin{abstract}
The suppression of high transverse momentum particles, recently discovered at RHIC, is commonly interpreted as due to parton energy loss. In high energy nuclear collisions, QCD jets would deposit a large fraction of
their energy and into the produced matter. The question of how this energy is degraded and whether
we can use this phenomenon to probe the properties of the 
produced matter is now under active discussion. It has been proposed
that if this matter, which is now being referred to as a 
{\em strongly coupled Quark-Gluon Plasma} (sQGP), may behave as a liquid 
 with a very small viscosity.
 In this case, 
 a very specific collective excitation should be produced,
called the ``conical flow'', similar e.g. to
the sonic booms generated by the shock waves produced by supersonic planes.
The RHIC experiments seem indeed to be obtaining some indication that the production of particles emitted opposite to a high-$p_t$ jet may actually be peaked away from the quenched jet direction, at an angle roughly consistent with the direction expected in case a shock wave is produced (i.e. orthogonal to the Mach cone).
In this note we speculate that for tagged heavy-quark jets one may observe a shrinkage of the Mach cone at moderate $p_t$.   
The experimental observation of such an effect would be a very good test for the validity of the whole picture currently emerging from the study of partonic matter in nuclear collisions.
\end{abstract}

\section{Introduction}
A strong suppression of the production of particles at high transverse momentum has been observed at RHIC. This is commonly interpreted as due to parton energy loss ({\em jet quenching}). See early papers \cite{early},
more recent developments \cite{Dok_etal,SZ_dedx} and a short summary
\cite{xnwang_workshop}.

\begin{figure}
 \includegraphics[width=7cm]{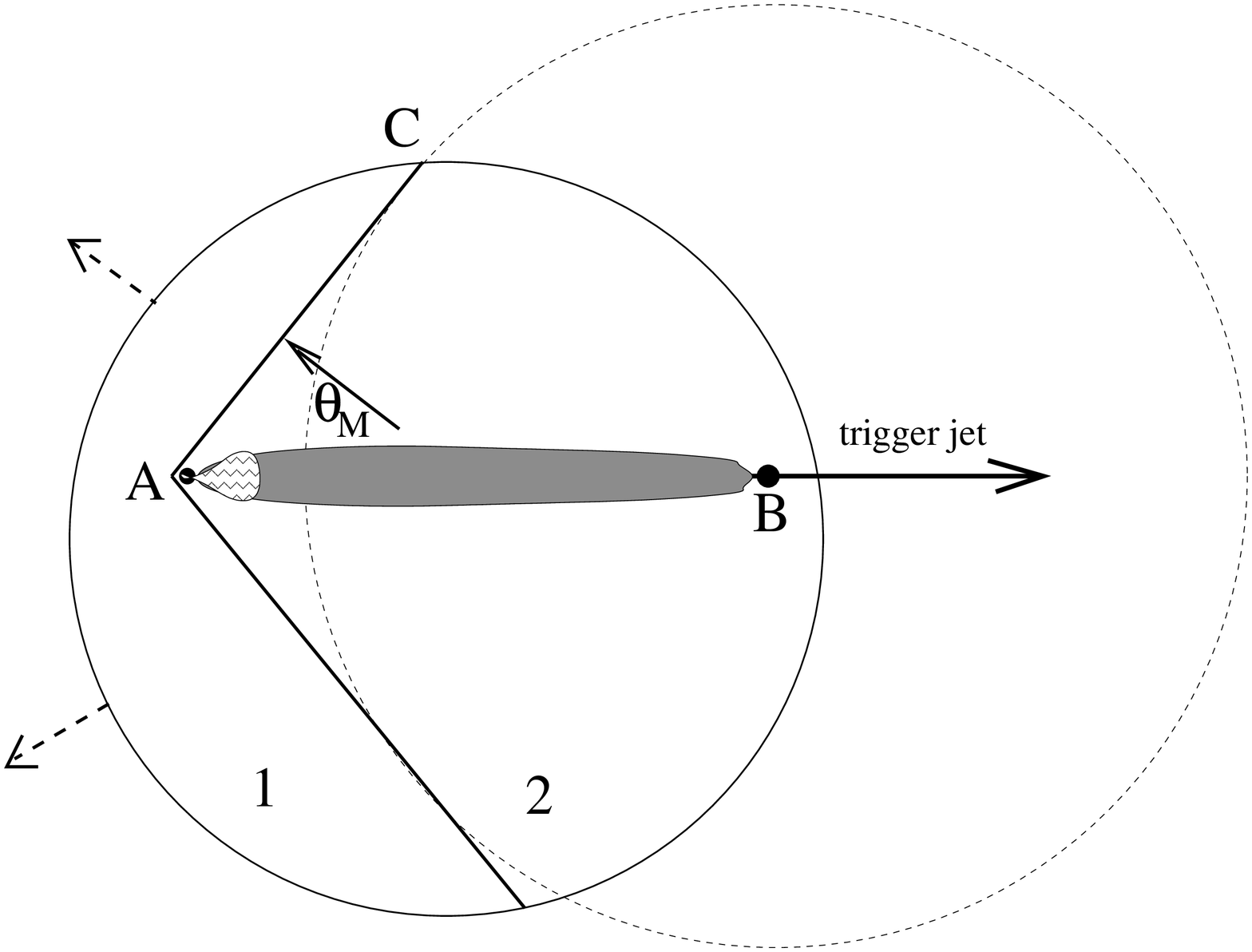}
 \includegraphics[width=7cm]{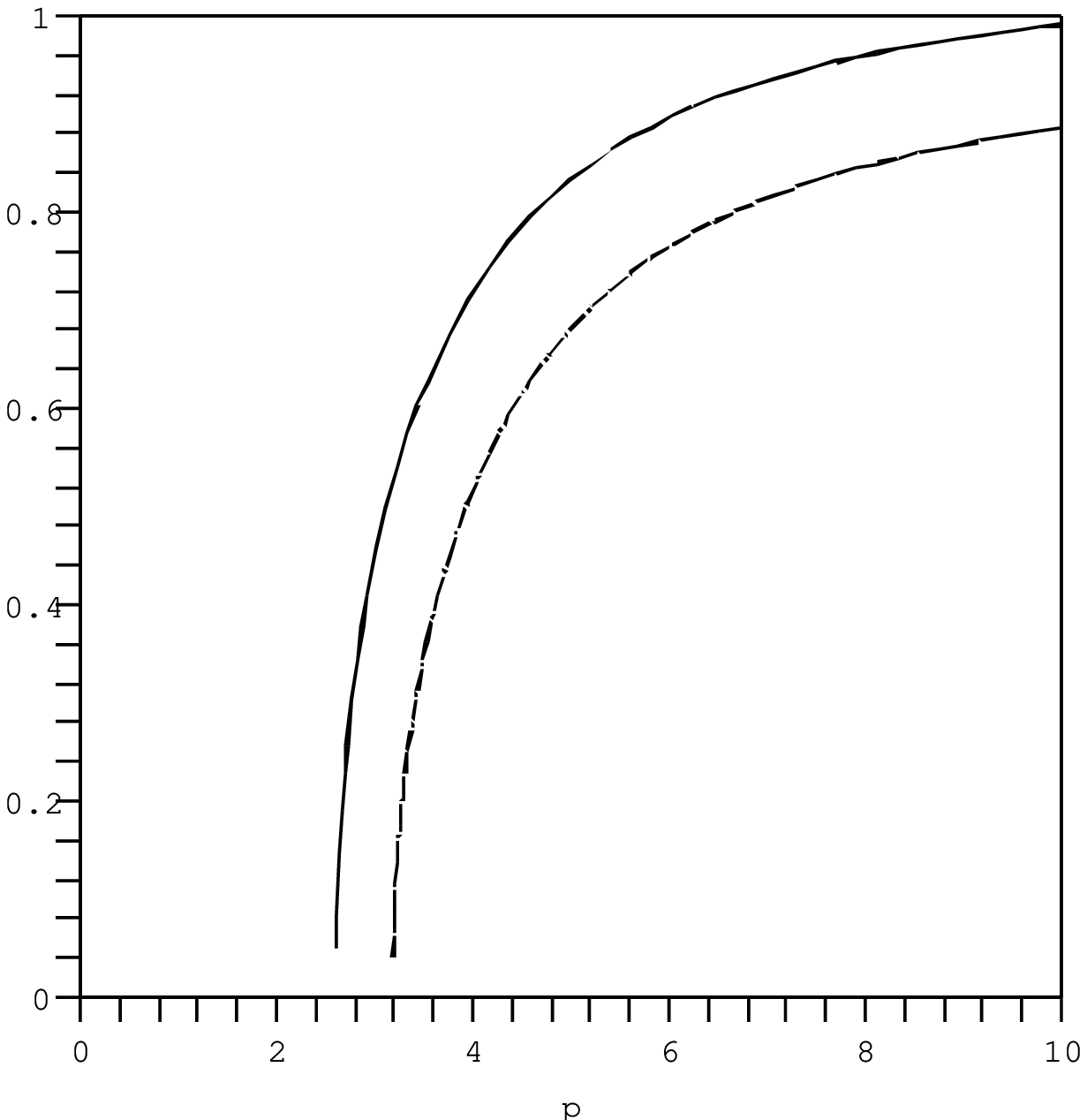}
 \caption{\label{fig_shocks}
(a) A schematic picture of flow created by a jet going through
the fireball. The trigger jet is going to the right
from the origination point (the black circle at point B)
from which sound waves start propagating as spherical waves (the
dashed circle). The
  companion quenched jet is moving to the left, heating the matter
and thus creating a cylinder of excited matter (shaded area).
 The head of the jet is a ``nonhydrodynamical core'' of the QCD gluonic shower,
formed by the original hard parton (black dot A).
The solid arrow shows a direction of flow normal to the shock cone at
the angle $\theta_M$, the 
dashed arrows show the direction of the flow after shocks hit the edge
of the fireball.
(b) Dependence of the Mach cone angle $\theta$ (rad) for LHC on the
momentum $p$ (GeV) of b-quark. The upper line is for $\bar
c_s^{LHC}=1/2$,
the lower for  $\bar
c=1/\sqrt{3}$.
}
 \end{figure}

An interesting question is whether the deposited energy may eventually be observed, and in what form.
Since the state produced at RHIC appears to be a liquid-like QGP \cite{sQGP}
with a very short dissipative length 
(small viscosity)
\cite{Derek_visc}, details such as the angular distribution of
emitted gluons may be rapidly forgotten and energy/momentum
be deposited locally. Further evolution would then be described
by relativistic hydrodynamics.
In \cite{CST} (in a brief form the idea was also mentioned in
\cite{Stocker})
 it was suggested 
 that such local deposition of energy and momentum could start a
 ``conical flow'' of matter. The arrangement is shown in Fig.\ref{fig_shocks}(a).

The solution of the linearized hydrodynamics equations \cite{CST} provides
a detailed picture of resulting flow of matter. 
 The main direction of flow
 is at the so called Mach angle relative to the jet velocity
\be 
\label{eqn_Mach}
cos\theta_M={\bar  c_s\over v_{jet}}
\ee
where $v_{jet}$ is the jet velocity and $\bar c_s$ is the (time averaged)
speed of sound in produced matter. 

For RHIC collisions the 
time-weighted  $\bar c_s$ till the freezeout time $\tau$ is estimated to be
 \be \bar c_s^{RHIC}={1/\tau}\int_o^\tau dt c_s(t)\approx .33 \ee
Since for light quark and gluon jets
$v_{jet}\approx c$,
 the conical flow was calculated in \cite{CST} 
to be at about $\theta=1.2$ radian
or at angles of about 70 degrees relative to the jet.
Experimental observations made at RHIC, first by the STAR collaboration
\cite{STAR_peaks} and then by PHENIX \cite{PHENIX_peaks}, indicate a
depletion of correlated particles in the direction of the quenched jet
and a peak with an angular position and shape in agreement
with hydro predictions. 

The study of such hydrodynamical phenomena like elliptic and conical flow
will also be one of the central issues for the LHC heavy-ion programme. 
For LHC $\bar c_s$  is expected to be larger, since the QGP phase is expected to be longer-lived: for estimates we will tentatively use 
$\bar c_s^{LHC}=1/2$. 
Furthermore, in order to start a process
of shock wave formation,   the jet velocity should be
higher than speed of sound at early times $v_{jet}>c_s(t\approx 0)$, 
for which we will use the value
for the QGP $c_s(0)=1/\sqrt(1/3)=.577$.

\section{Heavy Quark Shock Waves}

As discussed above, in the case of massless or very light partons 
(such as gluons or first-generation quarks), 
which travel at the speed of light, the expected angle of the conical
flow is fixed, as determined by the ratio of the speed of light to the
speed of sound in the medium.
 However, for massive partons of moderate momentum, the speed of
 propagation is substantially lower. If the picture outlined in the
 previous section is correct, this would then result in a
shrinking of
 the aperture of the shock-wave cone.

The case of the b quark is particularly interesting: with an estimated mass of the order of 4.5 GeV, the velocity of a b quark produced with a momentum of, say, 4 GeV is still expected to be only about 2/3 c. This suggests the possibility of actually observing the shrinking of the Mach cone in high energy heavy ion experiments, where heavy flavour jets are expected to be copiously produced.

For a massive parton of mass $m$ and momentum $p$, the Mach angle is given by:
\be \label{eqn MassiveMach}
cos\theta_M = {\bar c_s {\sqrt{p^2 + m^2}\over{p} }}
\ee
The result for the case of a b quark emitted at rapidity $y=0$ is
shown in Fig.1(b) as a function of the quark $p_t$ for two values of the speed of sound, using a value of 4.5 GeV for the mass of the b quark. 
Under such conditions, the b quark is subsonic for transverse
momenta of about 3 GeV, which should lead to disappearance of the
whole cone.
 At higher transverse momentum the aperture of the Mach cone
 approaches the asymptotic value (about 60 degrees for LHC) expected for massless partons, but for transverse momenta of the order of 3-4 GeV, the angle is expected to shrink to a value around 40 degrees. This may be sufficiently different from the asymptotic value to be distinguishable if one were able to obtain clean data.

\section{Experimental Considerations}

In Pb-Pb collisions at the LHC, from NTLO pQCD calculations and accounting for a ~ 15 \% shadowing effect, about 4.5 $b \bar{b}$ pairs are expected to be produced per central collision (\cite{Nicola-Andrea}). The Alice Collaboration, for instance, expects in its acceptance something of the order of one b-decay electron with a pT larger than 3 GeV every 50 central (5\% cross section) events, including tracking and electron identification inefficiencies (\cite{ALICE-PPR}). The possibility comes to mind of tagging a $b$ ($\bar{b}$) quark decay, and studying the bulk particle flow in the opposite hemisphere, where the companion $\bar{b}$ ($b$) quark is expected to be emitted, in the attempt to measure a possible deformation due to the presence of the shock wave produced by the propagating companion.

However, in order to attempt a correlation between a detected heavy quark decay and the bulk particle flow in the opposite hemisphere, one would have to start with a b-tagged sample of high purity. This requires some additional constraints. 

The ALICE, ATLAS and CMS LHC experiments all plan to deploy silicon vertex detectors which should provide for a resolution on the track impact parameter (in the transverse momentum range of a few GeV and above, relevant for the present discussion) of the order of tens of micrometers. With an average lifetime of beauty hadrons of the order of 500 $\mu$m, a selection cut on the impact parameter of the electron track relative to the primary interaction vertex, demanding a minimum separation of a few hundred $\mu$m should provide for an adequate sample purity (at the expense of a further reduction in the statistics). As an example, the ALICE Collaboration expects that an impact parameter cut of 200 $\mu$m on electrons with a pT larger than 3 GeV would provide a b decay sample with a purity of about 90\%, at the expense of an additional loss of about one order of magnitude in the statistics. This would still leave more than one such electron for every 500 events. For a sample size of $10^7$ central Pb-Pb events (which should be collectable in one month at the expected luminosity of $10^{27} cm^{-2}s^{-1}$), this would still yield a statistics of over $10^4$.
More serious problems would come from the fact that the correlation
between the transverse momentum of the electron and that of the parent
b hadron is smeared, the effect being worst for electrons originating
indirectly from b decays via the $b\rightarrow c$ decay chain.

A way to attack these effects, at the expense of a further loss in statistics, could be to require the electron to originate from a reconstructed displaced decay vertex. This would allow the reconstruction of a larger fraction of the b hadron invariant mass and a measurement of the original line of flight of the b hadron itself, and would therefore provide for a better measurement of the original b hadron $p_{T}$. In addition, a kinematical cut on the transverse momentum of the electron with respect to the reconstructed line of flight of the b hadron (requiring it to be larger than one-half of the mass of a charm hadron) would also allow to rejection of the background originating from the decay od prompt c hadrons, and to better control the smearing effects due to the $b \rightarrow c$ decay chain.

The possibility of implementing such a strategy in the environment of nuclear collisions has to be investigated.
Finally, although, as discussed above, there is some indication that the deformations due to shock-wave-induced conical flow may already be visible at RHIC, again the expected performances for such measurements of the planned LHC experiments in the environment of nuclear collisions, have yet to be estimated.

\section{Conclusions}
We have speculated on the possibility that the propagation of heavy quarks through the dense strongly interacting medium produced in high energy nuclear collisions may result in the production of shock waves of a reduced cone size with respect to those expected for light partons, and we have commented on the possibility of experimentally observing such an effect at the LHC.
The treatment presented in this letter is very crude: it only serves the purpose of estimating the magnitude of the effects. Also, the strongly interacting system produced in nuclear collisions at the LHC may be denser than that obtained at RHIC, resulting in a higher value for the speed of sound than that employed here, and therefore in a reduction of the momentum range over which the effect may be present.

Experimentally, the production rates should not by themselves be a showstopper, even if rather tight selection cuts are employed. The suggested effect, if existent, may however still in the end turn out to be too difficult to detect, due to the unavoidable real-life smearings and experimental biases. 
The capability of the individual LHC experiments in this respect has yet to be studied in detail with the inclusion of all relevant theoretical and experimental diluting effects. 

On the other hand, since heavy flavours are known to be produced in pairs in strong interactions and given that the production of heavy flavour particles in gluon jets is known to be negligible, the detection of a b decay is a sure tag for another b quark propagating roughly in the opposite direction. 
The detection of b quark shock waves with the properties discussed in this paper would allow us to exploit the properties of the dense medium produced in high energy nuclear collisions to observe, finally, the effects of the propagation of a single quark, and provide a very rough measurement of the propagating quark's inertial mass. We believe that such physics potential is enough to motivate further investigation in this direction.

\vskip 1cm
This work was partially supported by the Istituto Nazionale di Fisica Nucleare (Italy) and by the Department of Energy (U.S.A.) under grants DE-FG02-88ER40388
and DE-FG03-97ER4014.

\end{document}